\documentclass[aps,pra,reprint,twocolumn,superscriptaddress,floatfix]{revtex4}

\usepackage{amsmath}
\usepackage{graphicx}
\usepackage{dcolumn}
\usepackage{mathrsfs}
\usepackage{multirow}
\usepackage{mciteplus}
\usepackage{color}
\usepackage{epstopdf}

\begin{document}


\title{Optimized pulses for Raman excitation through the continuum: verification using multi-configurational time-dependent Hartree-Fock}


\author{Loren Greenman}
\email{lgreenman@lbl.gov}
  \affiliation{Chemical Sciences Division, Lawrence Berkeley National Laboratory, Berkeley CA 94720}
\affiliation{Department of Chemistry and Kenneth S. Pitzer Center for
  Theoretical Chemistry, University of California, Berkeley, CA 94720,
  USA} 
\author{K. Birgitta Whaley}
\affiliation{Department of Chemistry and Kenneth S. Pitzer Center for
  Theoretical Chemistry, University of California, Berkeley, CA 94720,
  USA} 
  \affiliation{Chemical Sciences Division, Lawrence Berkeley National Laboratory, Berkeley CA 94720}
\author{Daniel J. Haxton}
  \affiliation{Chemical Sciences Division, Lawrence Berkeley National Laboratory, Berkeley CA 94720} 
\author{C. William McCurdy}
  \affiliation{Chemical Sciences Division, Lawrence Berkeley National Laboratory, Berkeley CA 94720}
  \affiliation{Department of Chemistry, University of California, Davis, CA 95616 USA}
\date{\today}


\begin{abstract}
We have verified a mechanism for Raman excitation of atoms through continuum levels previously obtained by 
quantum optimal control using the multi-configurational time-dependent Hartree-Fock (MCTDHF) method.
For the optimal control, which requires running multiple propagations to
determine the optimal pulse sequence, we used the computationally
inexpensive time-dependent configuration interaction singles~(TDCIS) 
method. TDCIS captures all of the necessary correlation of the desired 
processes but assumes that ionization pathways reached via double 
excitations are not present.
MCTDHF includes these pathways and all multiparticle correlations in a 
set of time-dependent orbitals.
The mechanism that was determined to be optimal in the Raman excitation
of the Ne $1s^22s^22p^53p^1$ valence state via the metastable 
$1s^22s^12p^63p^1$ resonance state involves a sequential
resonance-valence excitation. First, a long pump pulse excites the
core-hole state, and then a shorter Stokes pulse transfers the
population to the valence state.
This process represents the first step in a multidimensional x-ray spectroscopy scheme that will provide a local probe
of valence electronic correlations.
Although at the optimal pulse intensities at the TDCIS level of theory the MCTDHF method predicts multiple ionization
of the atom, at slightly lower intensities (reduced by a factor of about 4) the TDCIS mechanism is shown to hold qualitatively.
Quantitatively, the MCTDHF populations are reduced from the TDCIS calculations by a factor of 4.
\end{abstract}

\maketitle

\section{Introduction}

Whereas linear spectroscopy directly measures the energies of states via the first-order response function,
multidimensional spectroscopy measures couplings between states using higher-order response functions.
Multidimensional spectroscopies are currently used to measure couplings in the regimes of radiowaves (NMR)~\cite{ernst1987principles, sattler1999heteronuclear, kanelis2001multidimensional},
infrared (vibrational)~\cite{larsen2001vibrations,khalil2003coherent}, and UV-Vis (photon echo)~\cite{mukamel2000mdfemto,jonas2003twodfemto,engel2007fmo,biggs2012twod,mukamel2013multidimensional}.
An x-ray analog of such spectroscopies could be used to measure couplings between localized core-hole excitations~\cite{tanaka2002coherent,mukamel2009coherent}.
Such couplings are due to valence electron interactions, and therefore x-ray multidimensional spectroscopy provides
a local probe of valence excitations.
However, complications arise due to the high energy of x-ray pulses, which can ionize samples or cause other unwanted processes to occur.

Multidimensional spectroscopy uses two or more frequencies to measure
the coupling between two excited states of an atom or molecule. In
multidimensional x-ray techniques, localized core-hole states can be 
addressed by one or more of these frequencies. An x-ray Raman
excitation of a valence excited state in a molecule can be correlated
with another core-hole excitation located far away to measure spatial
energy transfer~\cite{tanaka2002coherent}. Two Raman excitations can be 
used with a variable time delay to probe the relaxation of valence
excitons~\cite{zhang2014nonlinear}. In some schemes, the phases between
the different states are used to measure the coherence between two
excited states or to enhanced the signal considerably. It is therefore
important to ensure that the high energy of the x-rays does not ionize
the system or initiate spectator processes that reduce the coherence.

\begin{figure}
\includegraphics[width=0.7\linewidth]{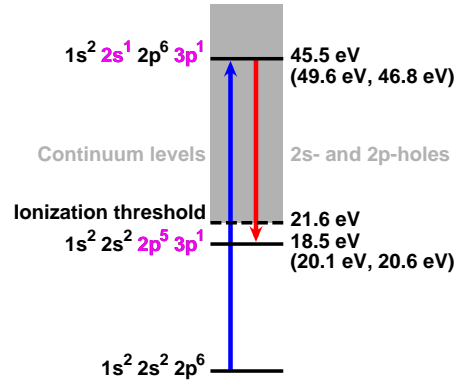}
\caption{(Color online)
The target Raman process is pictured. The pump (red) pulse excites the 
intermediate (2s-3p) state, then the Stokes (blue) pulse transfers the population 
to the desired (2p-3p) state.
The experimental energy levels are given along with the TDCIS (from diagonalization) 
and MCTDHF (determined as in Fig.~\ref{cwpulse_low_intensity}) energy levels in parentheses.
\label{scheme}}
\end{figure}

Two of us have recently obtained optimized pulses in a theoretical study
that perform the first crucial step of a multidimensional x-ray scheme while avoiding ionization~\cite{greenman2014laser}.
The design of the pulses in that study was accomplished by combining 
Krotov's optimal control method~\cite{Tannor92, SomloiCP93, konnov1999krotov,bartana2001krotov,JoseKochPRA08,reich2010monotonically} 
with time-dependent configuration interaction singles~(TDCIS) electronic 
dynamics including the ionization continuum~\cite{greenman2010implementation,xcid}.
The TDCIS method is a good choice for optimal control calculations, it is computationally cheap and captures low-order
electron correlation by including all singly excited electronic configurations.
However, TDCIS ignores multiply excited pathways, and so the reliability of the optimal pulses in an experimental setting is unclear.
These multiply excited pathways could lead to ionization, reducing the
overall yield of the final result. They could also reduce the coherence
predicted by TDCIS by interacting with the singly excited pathway and 
altering the phases of the states.
In this work, we use the multiconfigurational time-dependent Hartree-Fock (MCTDHF)~\cite{haxton2011multiconfiguration,haxton2012single}
method to verify that the pulses that were optimized using TDCIS to
 accomplish the population transfer shown in Fig.~\ref{scheme} in the
Ne~atom perform similar population transfers when the electronic 
dynamics are described with the inclusion of higher-order electron 
correlation.
MCTDHF includes all excitation pathways within a subset of orbitals, which are time-dependent (unlike TDCIS, which uses
time-independent orbitals).

This method simultaneously describes stable valence states, autoionizing 
states, and the photoionization continua, which are involved in these 
experiments, and this approach has been previously explored and 
developed by several groups \cite{alon2007unified,caillat2005correlated,kato2009timedependent,nest2007time,nest_lih2012, Miranda_2011, Madsen_2013,Sato_2013}. 
Briefly, our implementation solves the time-dependent Schr\"{o}dinger 
equation in full dimensionality, with all electrons active.  It 
rigorously treats the ionization continua for both single and multiple 
ionization using complex exterior scaling.   As more orbitals, and 
larger grids for describing them, are included, the MCTDHF wave function 
formally converges to the exact many-electron solution, but here the 
limits of computational practicality were reached with the inclusion of 
full configuration interaction with nine time dependent orbitals.  
While it is possible to do larger calculations on Ne using
MCTDHF~\cite{cryan2016optimizing} (we have used up to 14 orbitals),
we determined that these calculations required much smaller timesteps to
determine accurately the (small) populations of the states involved in
the Raman process. 
Nonetheless these calculations provide a substantial test of the 
assumptions of the simpler and more computationally tractable TDCIS 
approach.

Previously, we used MCTDHF to perform Raman excitation of atomic 
lithium~\cite{li2014population} and the NO molecule~\cite{haxton2014ultrafast}.
In both of those studies, as in Ref.~\cite{greenman2014laser} and also in the
current work, the first step in a multidimensional scheme (such as those
described above)
is attempted, and the intermediate state of the Raman process is a 
resonance state above the level of the electronic ionization continuum.
Additionally, all of these investigations have found adiabatic 
mechanisms such as stimulated Raman adiabatic passage (STIRAP)~\cite{bergmann1998stirap} 
to be ineffective at the energy and timescales of interest.
In our previous MCTDHF studies~\cite{li2014population,haxton2014ultrafast}, 
the potentially large amount of background ionization due to absorption 
of the x-ray pulses by spectator orbitals was avoided by choosing inner 
core levels to address with the pulses.
The high-energy x-rays that address these levels have a much lower cross section for absorption by spectator orbitals.
In the case of of our previous study of Raman population transfer in the 
Li atom~\cite{li2014population}, there were no occupied $p$-orbitals to 
contribute to background ionization.
In contrast, the previous study for neon~\cite{greenman2014laser}
employed optimal control theory to find pulses that minimize background 
ionization but penalizes distance from some guess pulse.
A mechanism was thereby found to excite a Raman excitation using pulses 
with lower energies, although a smaller fraction of the final 
wavefunction is in the Raman state.
That study also considered coherent excitation of the Raman state~\cite{greenman2014laser}, 
and optimal pulses were also obtained that 
excite the Raman state with a fixed phase relative to the ground state.

These optimizations performed with TDCIS produced specific pulses but also
revealed a more general mechanism for generating pulse
sequences that perform x-ray Raman while avoiding ionization~\cite{greenman2014laser}.
In this sequential mechanism, a long pump pulse is first used to selectively excite population from the ground state 
to the intermediate state, and then a shorter Stokes pulse is used to transfer population from the intermediate state to the 
desired state.
The long pump pulse selects the transition to the intermediate state, which is located close in energy to
a dense number of continuum states, and avoids background transitions to those states.
If a specific phase is desired between the Raman state and ground state, it can be imprinted via the carrier envelope
phase of the pump pulse~\cite{greenman2014laser}.
The length of the Stokes pulse is somewhat flexible, but it must be short enough to overcome autoionization 
from the intermediate state.
The ideal placement of the Stokes pulse is near the peak of the intermediate state population, 
which TDCIS predicts to be slightly before the pump pulse maximum for a pump pulse on the order of 50 fs.

Here, as before~\cite{greenman2014laser}, we use Ne as an example 
because of its accessibility to tabletop experiments through the rapidly 
advancing availability of XUV high harmonic generation
and free electron lasers such as FERMI@Elettra~\cite{fermifel}. 
The levels we are targeting are shown in Fig.~\ref{scheme}.
The intermediate state is the 2s-3p state of Ne, which lies above the ionization threshold.
The target state is the 2p-3p valence excitation.

We find that up to a factor less than an order of magnitude, electron 
correlation effects captured using MCTDHF do not destroy the efficacy of the optimal pulses. 
However, this is true only up to a certain intensity, above which 
multiple ionization pathways make TDCIS unreliable.

\section{Theory}

Both the time-dependent configuration interaction singles (TDCIS)~\cite{greenman2010implementation, xcid} method and the multiconfigurational time-dependent Hartree-Fock (MCTDHF)~\cite{haxton2011multiconfiguration,haxton2012single} method
choose a reference configuration ($\vert \Phi _0\rangle$) that is an antisymmetrized product of $N_e$ single-particle orbitals,
\begin{equation}
\vert \Phi _0\rangle = \vert \phi _1 \phi _2 \ldots \phi _{N_e}\rangle .
\label{reference}
\end{equation}
Both methods describe the many-electron wavefunction using this reference and configurations obtained by exciting particles from the reference,
\begin{eqnarray}
\label{singles}
\vert \Phi _i^a\rangle &=& \hat{a}^\dagger_a \hat{a}_i \vert \Phi _0\rangle \\
\label{doubles}
\vert \Phi _{i,j}^{a,b}\rangle &=& \hat{a}^\dagger_a \hat{a}^\dagger_b \hat{a}_j\hat{a}_i \vert\Phi _0\rangle , \ldots,
\end{eqnarray}
where $i,j$ denote orbitals occupied in the reference and $a,b$ denote unoccupied orbitals
and $\hat{a}$ and $\hat{a}^\dagger$ denote annihilation and creation operators, respectively.

In the configuration interaction singles (CIS) method, the reference (Eq.~\eqref{reference}) and \textit{all} singly-excited configurations (Eq.~\eqref{singles}) are included 
(up to a very high energy cutoff),
\begin{equation}
\vert \Psi (t)\rangle = \alpha _0 (t) \vert\Phi _0\rangle + \sum _{i,a} \alpha _i^a(t) \vert\Phi _i^a\rangle .
\label{tdciswf}
\end{equation}
In this configuration space, dynamic electron correlation between singly-excited configurations is taken into account.
Due to Brillouin's theorem, there is no mixing between the reference configuration and excited configurations due to Coulomb interactions.
CIS therefore provides a first-order description of excited states dominated by single-particle configurations.
Excitations that involve multiple occupied orbitals are not qualitatively well-described by CIS.
Time-dependent CIS (TDCIS) uses time-dependent coefficients on the CIS configurations to describe the time-evolving wavefunction.
The orbitals $\phi_i$ remain time-\textit{independent}, in contrast to the MCTDHF method.
TDCIS can not describe multiple ionization pathways.

The MCTDHF method~\cite{caillat2005correlated,alon2007unified,nest2007time,nest2008pump,kato2009timedependent,haxton2011multiconfiguration,haxton2012single}, as implemented in Refs.~\cite{haxton2011multiconfiguration,haxton2012single}, uses a smaller subset of $N_o$ orbitals, $\{\phi _{sub}\} = \{ \phi_1, \ldots, \phi_{N_o}\}$, but includes all configurations in this subset.
This means that multiply ionized pathways can now be described.
The coefficients on each configuration and the shape of the orbitals that define the reference and excited configurations are both time-dependent.
\begin{eqnarray}
\nonumber
\vert\Psi (t)\rangle &=& \alpha _0 (t) \vert\Phi _0(t)\rangle + \sum _{i,a \in \{\phi _{sub}\}} \alpha _i^a(t) \vert\Phi _i^a(t)\rangle \\
&&+ \sum _{i,j,a,b \in \{\phi _{sub}\}} \alpha _{i,j}^{a,b}(t) \vert\Phi _{i,j}^{a,b}(t)\rangle + \ldots
\label{mctdhfwf}
\end{eqnarray}
It should be noted that an implementation using finite element DVR grids
of MCTDHF using restricted configuration spaces has also been developed~\cite{haxton2015two}.
MCTDHF in a small (practical) space of orbitals mainly captures static
(nondynamic) correlation, i.e., the contribution from configurations that at zeroth order define the wavefunction.
For instance, double ionization from the core is described at zeroth
order using a doubly excited configuration; TDCIS can not describe this.
While reducing the number of orbitals leads to a greater amount of dynamic correlation being left out, the time-dependent nature of the orbitals 
could possibly reintroduce some dynamic correlation back into the calculation.
Furthermore, dynamic correlation tends to lead only to quantitative, and not qualitative, corrections to the wavefunction.

One further difference between the TDCIS method employed in Ref.~\cite{greenman2014laser} and the MCTDHF method is the description of the ionization continuum.
The TDCIS method uses a complex absorbing potential (CAP)~\cite{goldberg1978cap,santra2002cap,greenman2010implementation}, an imaginary quadratic potential that is turned on after a cutoff radius.
CAPs can be tuned to capture a small number of resonance energies correctly, but they can also perturb the bound states
and continuum states outside the region for which they're tuned.
MCTDHF instead uses exterior complex scaling (ECS)~\cite{simon1979definition,mccurdy2004solving}, in which the spatial coordinates are scaled into the complex plane by an angle $\theta$.
Grid implementations of ECS have been shown to effectively treat 
single and double ionization continua~\cite{mccurdy2004solving}, and they do not perturb the 
bound states.

Comparing MCTDHF and TDCIS propagations using the optimal pulses
previously determined with Krotov's method~\cite{greenman2014laser}, 
thus provides a fuller view of the time-dependent processes in the x-ray 
Raman excitation of atoms.

The MCTDHF calculations presented here were obtained using a space 
of 9 time-dependent orbitals, the $1s$, $2s$, $2p$, $3s$, and 
$3p$~orbitals of Ne.
There are 4116 configuration state functions in this active space.
The orbitals are described with a finite element version of the discrete 
variable representation (DVR) in the radial degree of freedom, a DVR in 
the polar angle, $\theta$, and analytical functions of the azimuthal 
angle, $\exp(i m \varphi)$.
We used a radial grid of six 9.0~Bohr~elements, each with 19 grid 
points, and an angular grid of 5$^\mathrm{th}$ order in $\theta$ and analytic 
functions capable of describing  angular momentum states of up to $m=2$.
The last element was complex scaled using an angle of 0.4 radians.

\section{Results and discussion}

The mechanism for x-ray Raman excitation of atoms while avoiding ionization is as follows:
first, a long pump pulse is used to selectively excite the intermediate state (the 2s-3p state of Ne), 
followed by a shorter Stokes pulse that beats the autoionization of the intermediate state and 
transfers the population to the desired state (in Ne, the 2p-3p state).
This mechanism was discovered in Ref.~\cite{greenman2014laser} using TDCIS and optimal control theory, 
and it was used there to develop experimentally realizeable pulses.
A 50$\,$fs, 71$\,\mu$J~pump pulse and 0.5$\,$fs, 0.71$\,\mu$J~Stokes
pulse represent one choice of pulses. 
Other options were presented in Ref.~\cite{greenman2014laser}, using
variable lengths of the Stokes pulse.
The peak intensity of the pump pulse was $6.1 \times 10^{14}\,$W/cm$^2$. 

Since MCTDHF and TDCIS have different descriptions of the electron correlation, 
the transition frequencies at each level of theory will be different.
Therefore, the MCTDHF transition frequencies must first be obtained. We accomplished
this by running various continuous wave (CW) pulses with many central frequencies.
The relevant MCTDHF frequencies can also be determined by Fourier
transforming the dipole after exciting with a short pulse, but this
requires long propagations after the pulse and it is also difficult to
get some resonance states in this manner.

\begin{figure}
\includegraphics[width=0.9\linewidth]{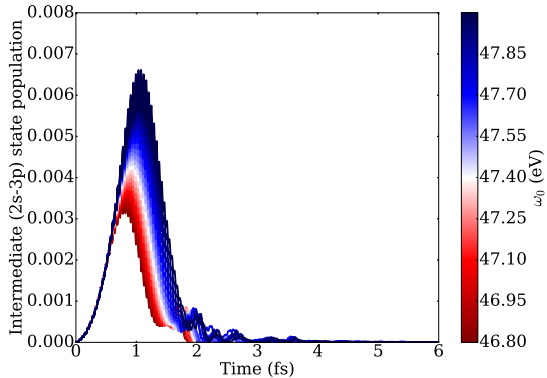}
\caption{(Color online)
MCTDHF intermediate state (2s-3p) populations for CW pulses at intensities optimized using TDCIS.
Pulses are shown for a number of different central frequencies $\omega _0$ (see colorbar)
The opening of double and higher ionization channels imposes an intensity limit on the pulses.
The optimal intensity at the TDCIS level of theory is above this limit, 
which leads to ionization rather than populating the intermediate state.
\label{cwpulse_high_intensity}}
\end{figure}

Fig.~\ref{cwpulse_high_intensity} shows the results of one such set of computations with the peak intensities from the TDCIS optimal pulses.
The intermediate state populations are shown, with colors ranging from red to blue for central frequencies from 46.8 to 48.0 eV.
The optimal TDCIS intermediate state populations reached around 0.08, 
but the MCTDHF populations in Fig.~\ref{cwpulse_high_intensity} are much lower, less than 0.01.
At these intensities, the MCTDHF and TDCIS results differ significantly, and the most likely explanation is that
multiply ionized pathways are important at these intensities.
TDCIS does not take these pathways into account.
At an intensity of $6 \times 10^{14}\,$W/cm$^2$, approximately 3.5 photons/fs cross the atomic radius of Ne, 
which could lead to the absorption of 2 or more photons and ionize the atom.
A reduction of the intensity by a factor of about 4, however, returns
the system to the single-excitation regime which TDCIS describes well.

\begin{figure*}
\includegraphics[width=0.95\linewidth]{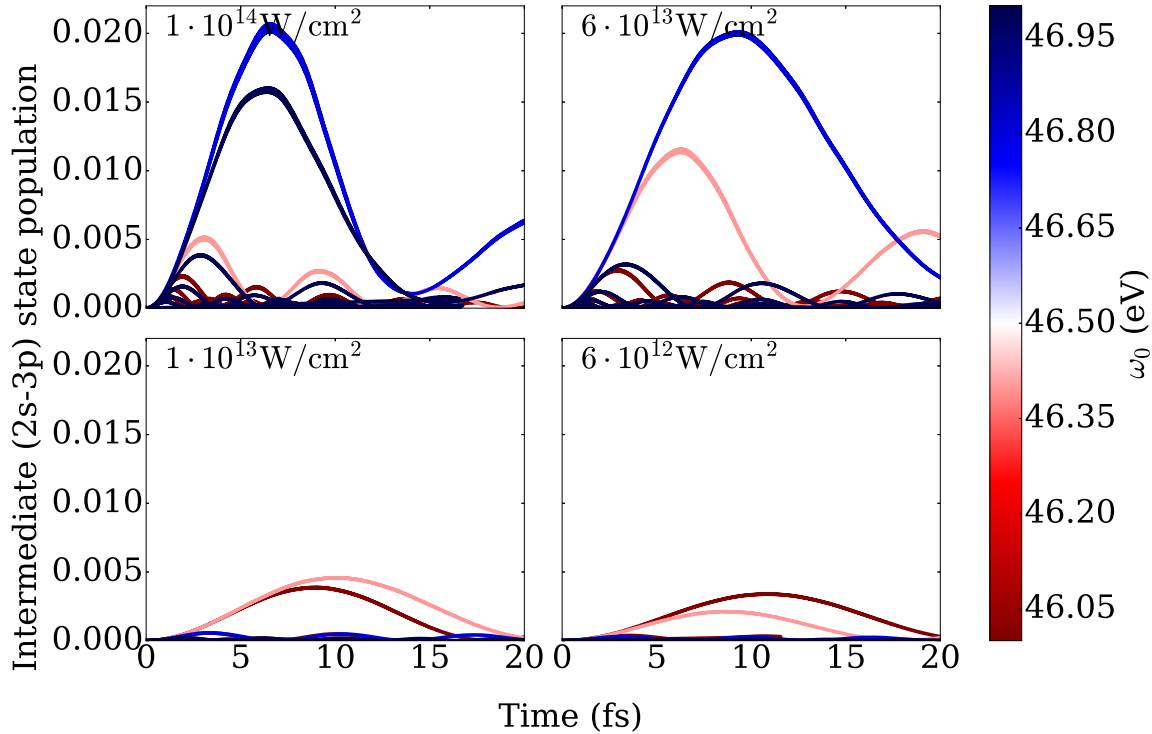}
\caption{(Color online)
Intermediate state (2s-3p) populations are shown at the MCTDHF level of theory for intensities lower than 
the optimal TDCIS intensity.
The multiple ionization channels are now closed, and the order-of-magnitude of the TDCIS and MCTDHF 
intermediate state populations are now the same.
The optimal central frequency for intermediate state population redshifts as the intensity is lowered.
At an intensity of $1 \cdot 10^{14}\,$W/cm$^2$
, the optimal central frequency of the pump pulse is 46.8 eV.
\label{cwpulse_low_intensity}}
\end{figure*}

In Fig.~\ref{cwpulse_low_intensity}, the lower intensity regimes are shown.
Intermediate state populations for peak intensities of $10^{14}$, 1 and 5 $\cdot 10^{13}$,
and $5 \cdot 10^{12}\,$W/cm$^2$~are shown.
At these intensities, the intermediate state is populated at the same order of magnitude
as estimated by TDCIS at the same intensities.
As expected from TDCIS, the higher intensities populate the intermediate state more
(as long as the multi-ionization threshold is avoided).
At $10^{14}$ and $5 \cdot 10^{13}\,$W/cm$^2$, intermediate state populations of about 0.02
are reached.
This is a factor of 4 lower than the TDCIS result.
For both of these intensities, the optimal pump pulse central frequency is found to be 46.8 eV.
At the lower intensities, the intermediate state is not populated very much.
This is also found at the TDCIS level of theory.
As the intensity of the pump pulse is lowered, the optimal central frequency is redshifted.

\begin{figure*}
\begin{minipage}[t]{0.49\linewidth}
\includegraphics[width=0.95\linewidth]{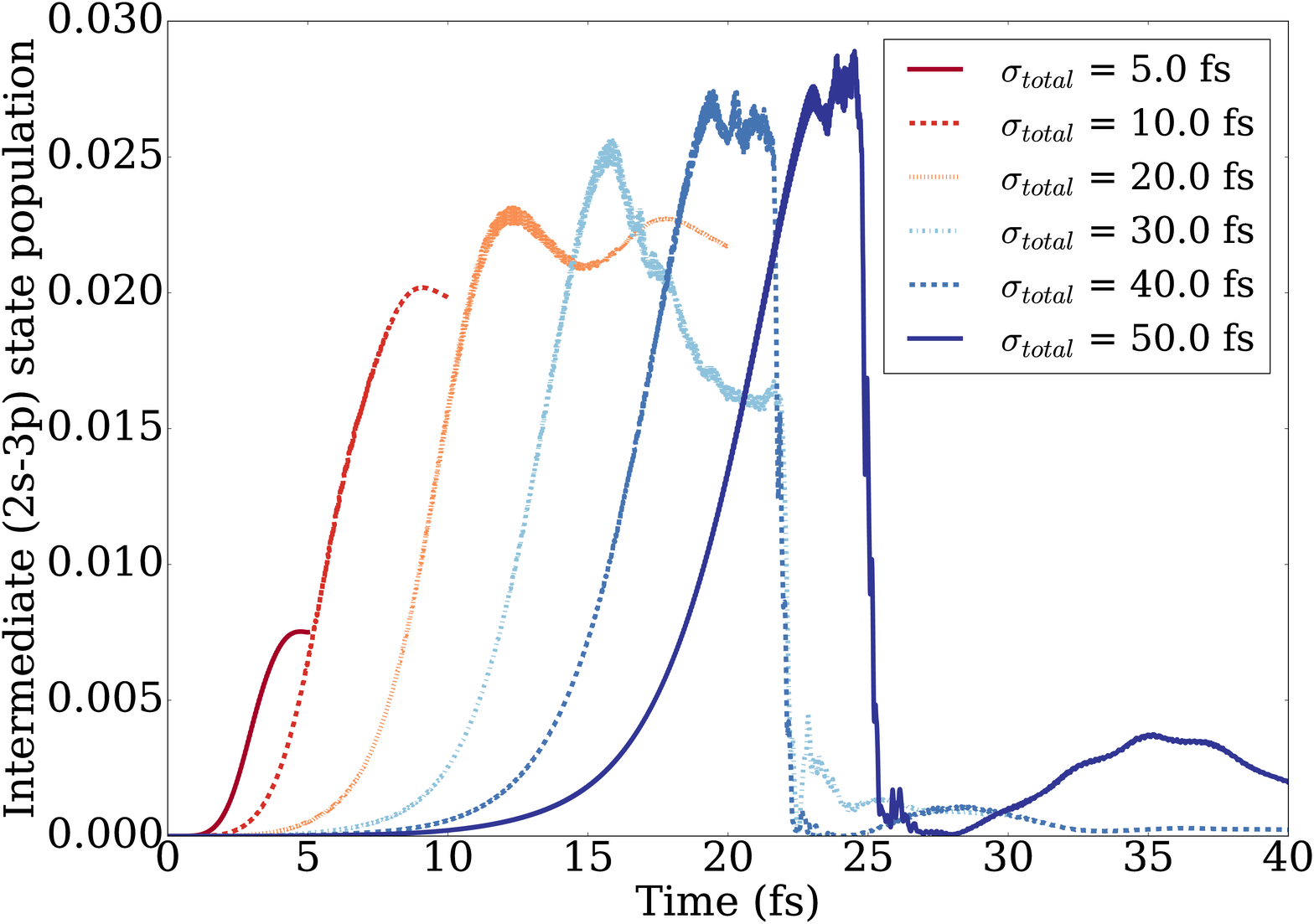}
\caption{(Color online)
The intermediate state (2s-3p) population during the pulse is shown for increasing pulse lengths $\sigma _{total}$, 
where the pulse envelope is given by $\mathcal{E}(t) = \mathcal{E}_0 \sin ^2 (\Omega t)$ and $\sigma _{total} = \pi /  \Omega$.
The maximum intermediate state population increases with pulse length, with the increases slowing as the pulse length grows
at an intensity of $10^{14}\,$W/cm$^2$.
The TDCIS optimal strategy is to maximize the intermediate state population and then use the Stokes pulse to 
transfer the intermediate state population to the desired state, and the maximum intermediate state population
reachable is around 0.03.
Longer pump pulses seem also to induce multiple ionization around the peak of the pulse.
\label{pulse_length_comparison}}
\end{minipage} \hfill
\begin{minipage}[t]{0.49\linewidth}
\includegraphics[width=0.95\linewidth]{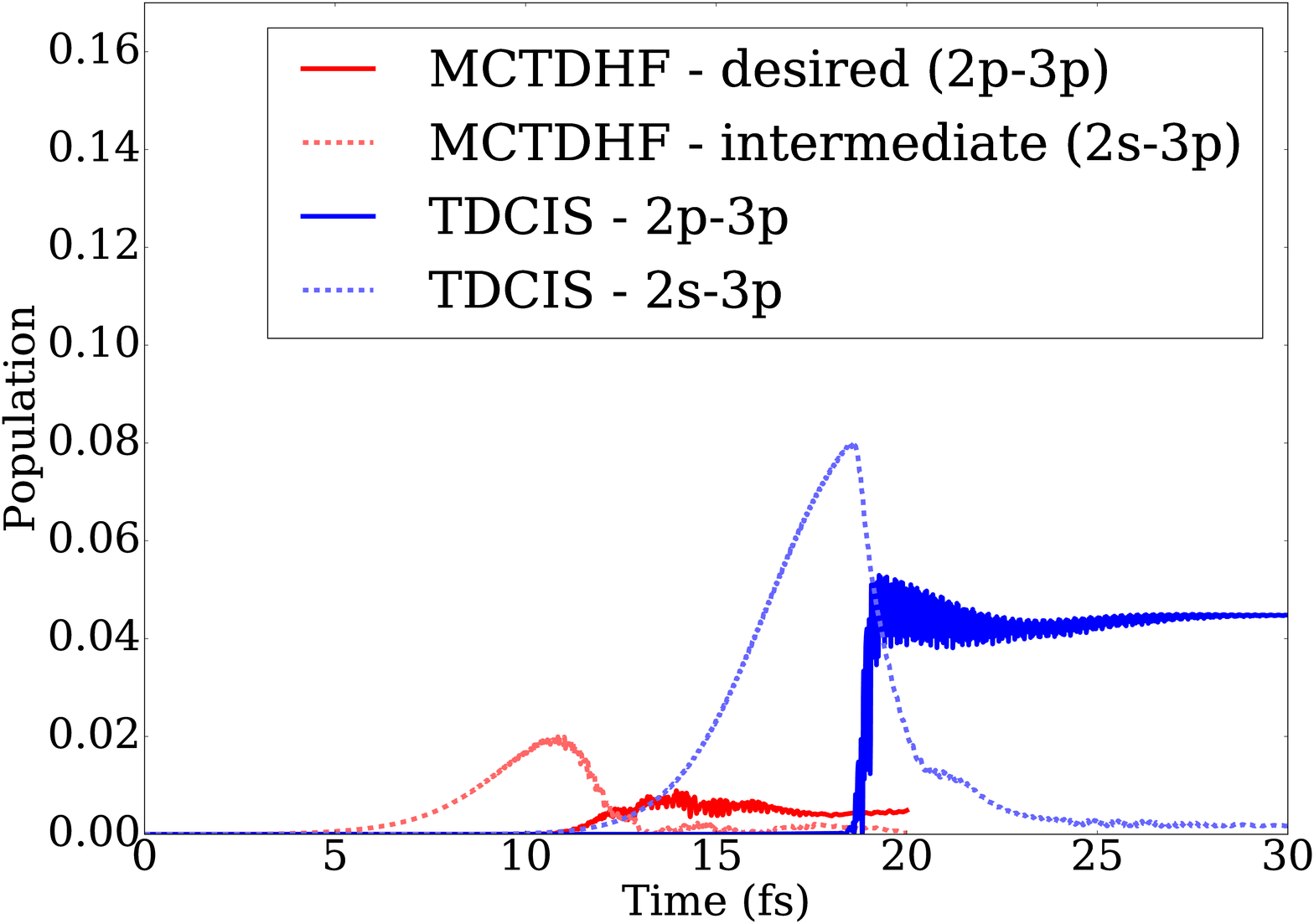}
\caption{(Color online)
The optimal TDCIS pulse (populations in blue) is compared with a similar (shorter) MCTDHF (populations in red) pulse with the same time ordering.
The desired state (dark lines) and intermediate states (light lines) are shown.
The qualititative features of the TDCIS and MCTDHF results are the same.
The MCTDHF populations are smaller by a factor of about 4. 
\label{method_comparison}}
\end{minipage}
\end{figure*}

With the peak intensity and central frequency for the pump pulse fixed at the 
values determined using the CW pulses, we test the effect of increasing the 
duration $\pi/\Omega$ of pulses shaped using a $\sin ^2(\Omega t)$ function. The
results are shown in Fig.~\ref{pulse_length_comparison}.
Similar to what was seen in the TDCIS results~\cite{greenman2014laser}, increasing the 
pump pulse length is found to be generally favorable.
The maximum intermediate state population increases largely at first, and then 
slightly as the pulse is made longer.
An intermediate state population of about 0.03 can be reached using a 50 fs pump pulse, 
but at these pulse durations it again appears multiple ionization pathways start to interfere.
For the 30, 40, and 50 fs pulse durations a dip in the population can be seen that
suggests that higher-order effects are beginning to occur.
Since the maximum intermediate state population increases only slightly above 20 fs,
and there are no observable multiple ionization effects at this pulse duration,
we use the 20 fs pump pulse when determining the optimal Stokes pulse parameters 
at the MCTDHF level.

Using the same method of determing the optimal central frequency and peak intensity of the Stokes pulse
with CW pulses, we determined that the intensity of the Stokes pulse predicted by TDCIS does not
introduce multiple ionization pathways.
Additionally, a number of calculations were run to determine the optimal central time of the Stokes pulse.
The resulting set of pump and Stokes pulses were used to determine the populations of the 
intermediate 2s-3p and desired 2p-3p states for Raman excitation of Ne, and compared with the optimal TDCIS 
pulse set in Fig.~\ref{method_comparison}.

Qualitatively, the TDCIS and MCTDHF optimal pulses are very similar.
A simple sequential population of the intermediate state followed by population transfer to the desired state
can be seen.
At the TDCIS level of theory, the intermediate state is populated to a level of 0.08, and about half of this
population can be transferred to the desired state.
We found in Ref.~\cite{greenman2014laser} that coupling between excitation channels induced by electron correlation 
keeps the entire population of the intermediate state from being transferred to the desired state.
At the MCTDHF level of theory, we have already determined that the intermediate state can be 
populated to a level of 0.02, and this can again be seen in Fig.~\ref{method_comparison}.


\section{Conclusion}

We have used the multiconfigurational time-dependent Hartree-Fock (MCTDHF) method in order to verify the performance of 
optimal pulses for x-ray Raman excitation of atoms.
This excitation represents the first step towards multidimensional x-ray spectroscopy, a tool for the direct and local measurement of electronic
interactions in valence levels.
The pulses were previously obtained in Ref.~\cite{greenman2014laser} using quantum optimal control theory combined with the time-dependent configuration interaction singles (TDCIS) method.
MCTDHF includes multiple excitation pathways that TDCIS does not, and
some of these were found to be important in the current study.
While some care is required to avoid such pathways when using TDCIS, the
qualitative features of the processes predicted by TDCIS were
nevertheless found
to extend to the more detailed calculations.
TDCIS, therefore, is an appropriate tool for optimal control calculations, having the advantages of speed while not sacrificing qualitative accuracy.

Using the combined Krotov optimal control and TDCIS method, we had previously determined a mechanism for avoiding ionization 
while performing the x-ray Raman excitation of atoms~\cite{greenman2014laser}.
First, the intermediate state is excited using a long pump pulse to selectively address the frequency of the desired transition.
Then, a short Stokes pulse is applied near the maximum intermediate state population to drive population to the desired valence state.
This pulse sequence avoids ionization, which is mainly due to direct ionization of the spectator orbitals (the 2p orbitals in the case of Ne).
This general scheme is supported by the MCTDHF calculations, however some details of its implementation differ from TDCIS.
At the intensities that are found to be optimal using TDCIS, multiple ionization pathways are found to occur using MCTDHF.
These processes dominate and very little population can be transferred to the intermediate state.
At slightly lower intensities, the mechanism found using TDCIS is again qualitatively successful.
Quantitatively, a factor of about 4 differentiates the TDCIS and MCTDHF populations.
This factor is likely due to the competing multiply-excited pathways
that are not present in TDCIS.

Using TDCIS, we determined that x-ray Raman excitation of Ne was experimentally feasible at the free electron laser facility FERMI@Elettra~\cite{fermifel}.
The pulses we have now found to be succesful at the MCTDHF level of
theory are also possible at that facility.
Specifically, a pump pulse with a duration of $20\,$fs and power of $0.6\,\mu J$ can be used
to Raman excite Ne and avoid ionization.

\begin{acknowledgments}
Work performed at Lawrence Berkeley National Laboratory was supported by
the US Department of Energy Office of Basic Energy Sciences, Division of
Chemical Sciences Contract  DE-AC02-05CH11231 and made use of the
resources of the National Energy Research Scientific Computing Center, 
a DOE Office of Science User Facility.   
KBW was supported through the Scientific Discovery
through Advanced Computing (SciDAC) program funded by the U.S.
Department of Energy, Office of Science, Advanced Scientific Computing
Research, and Basic Energy Sciences.
We also acknowledge funding through the US DOE Early
Career program and the Peder Sather Center.
\end{acknowledgments}
\bibliography{UltrafastControl}

\end{document}